\documentclass[11pt,twocolumn]{article}

\usepackage[USenglish]{babel} %francais, polish, spanish, ...
\usepackage[T1]{fontenc}

\usepackage{lmodern} %Type1-font for non-english texts and characters
\usepackage{setspace}
\usepackage{bookmark}
\usepackage{amsmath}
\usepackage{amsthm}
\usepackage{amsfonts}
\usepackage{floatrow}
\usepackage{fancyhdr}
\usepackage{graphics}
\usepackage{epsfig}
\usepackage{graphicx}
\usepackage{textcomp}
\usepackage{amsmath}
\usepackage{amsfonts}
\usepackage{amssymb}
\usepackage{mathrsfs}
\usepackage{enumerate} 
\usepackage{epstopdf} 
\usepackage{float}
\usepackage{subcaption}
\usepackage{gensymb}
\usepackage{cleveref}
\usepackage{placeins} 
\usepackage{sidecap}
\usepackage{flafter}  
\usepackage{setspace}
\usepackage[font=footnotesize,labelfont=bf]{caption}
\usepackage[left=2cm,right=2cm,top=2cm,bottom=2cm]{geometry}
\sidecaptionvpos{figure}{t}
\begin{document}
\setlength{\parskip}{0pt}
\title{Emergence of Active Nematics in Bacterial Biofilms }
\author{ Yusuf Ilker Yaman$^1$, Esin Demir$^2$, Roman Vetter$^3$, Askin Kocabas$^{1,4}$\footnotemark  \\ \small $^1$Department of Physics, Ko\c{c} University, 34450 Sar{\i}yer, Istanbul, Turkey\\ \small $^2$Bio-Medical Sciences and Engineering Program, Ko\c{c} University, 34450 Sar{\i}yer, Istanbul, Turkey\\ \small$^3$Department of Biosystems Science and Engineering, ETH Zurich, 4058 Basel, Switzerland\\ \small$^4$Ko\c{c} University Surface Science and Technology Center, Ko\c{c} University, 34450 Sar{\i}yer, Istanbul, Turkey
}

\date{\vspace{-5ex}}
%\date{\today}

\twocolumn[
\begin{@twocolumnfalse}
\maketitle
\begin{abstract}
\noindent
Growing tissue and bacterial colonies are active matter systems where cell divisions and cellular motion generate active stress. Although they operate in the non-equilibrium regime, these biological systems can form large-scale ordered structures such as nematically aligned cells, topological defects, and fingerings. Mechanical instabilities also play an essential role during growth by generating large structural folding. How active matter dynamics and mechanical instabilities together develop large-scale order in growing tissue is not well understood. Here, we use chain forming \textit{Bacillus subtilis}, also known as a biofilm, to study the direct relation between active stress and nematic ordering. We find that a bacterial biofilm has intrinsic length scales above which series of mechanical instabilities occur. Localized stress and friction control both linear buckling and edge instabilities. Remarkably, these instabilities develop nematically aligned cellular structures and create pairs of motile and stationary topological defects. We also observe that stress distribution across the biofilm strongly depends on the defect dynamics which can further initiate the formation of sporulation sites by creating three-dimensional structures. By investigating the development of bacterial biofilms and their mechanical instabilities we are proposing a new type of active matter system which provides a unique platform to study the essential roles of nematics in growing biological tissue.
\end{abstract}

\end{@twocolumnfalse}
]
	
\renewcommand{\thefootnote}{\fnsymbol{footnote}}
\footnotetext{$^*$Corresponding author: akocabas@ku.edu.tr}
\section*{Main}
Biofilm formation is a collective response of bacteria$^{1-6}$. Depending on the availability of food and environmental conditions$^7$, \textit{B. subtilis} produce matrix proteins and initiate the formation of a biofilm$^2$. During biofilm development, motile bacteria differentiate into an aligned chain of cells. Growing chains further develop fibers and bundles which shape the overall biofilm morphology$^{8-13}$. These distinct aligned structures promote sliding of a colony on a solid surface where the swimming behavior is not efficient$^{12,14}$.

Aligned cellular structures are also observed in a variety of biological phenomena. During wound healing, migrating cells align and form fingering structures at the leading edge of the tissue$^{15}$. Similarly, cultured cells$^{16-18}$ and isolated bacterial colonies$^{19-23}$ can form nematic alignment and modulate the cellular density and active stress. 

Recent studies have shown that liquid crystal theory can provide a suitable framework to study the dynamics of growing tissue as an active nematic system. Mainly, the dynamics of cellular alignment, topological defects, and edge instabilities have been explored$^{16-20}$. Here, by investigating the formation of a bacterial biofilm starting from a single bacterium, we have studied the detailed mechanical instabilities driving the dynamics of nematic active matter. We revealed the direct relation between local stress and localized buckling. This system provides not only a new platform to observe the essential dynamics of active nematics in a growing tissue but also the mechanical insights to explore the physics of nonbiological active nematics (AN)$^{24-28}$.

\section*{Results}
\begin{figure*}[t]
\centering
		\includegraphics[width=\linewidth]{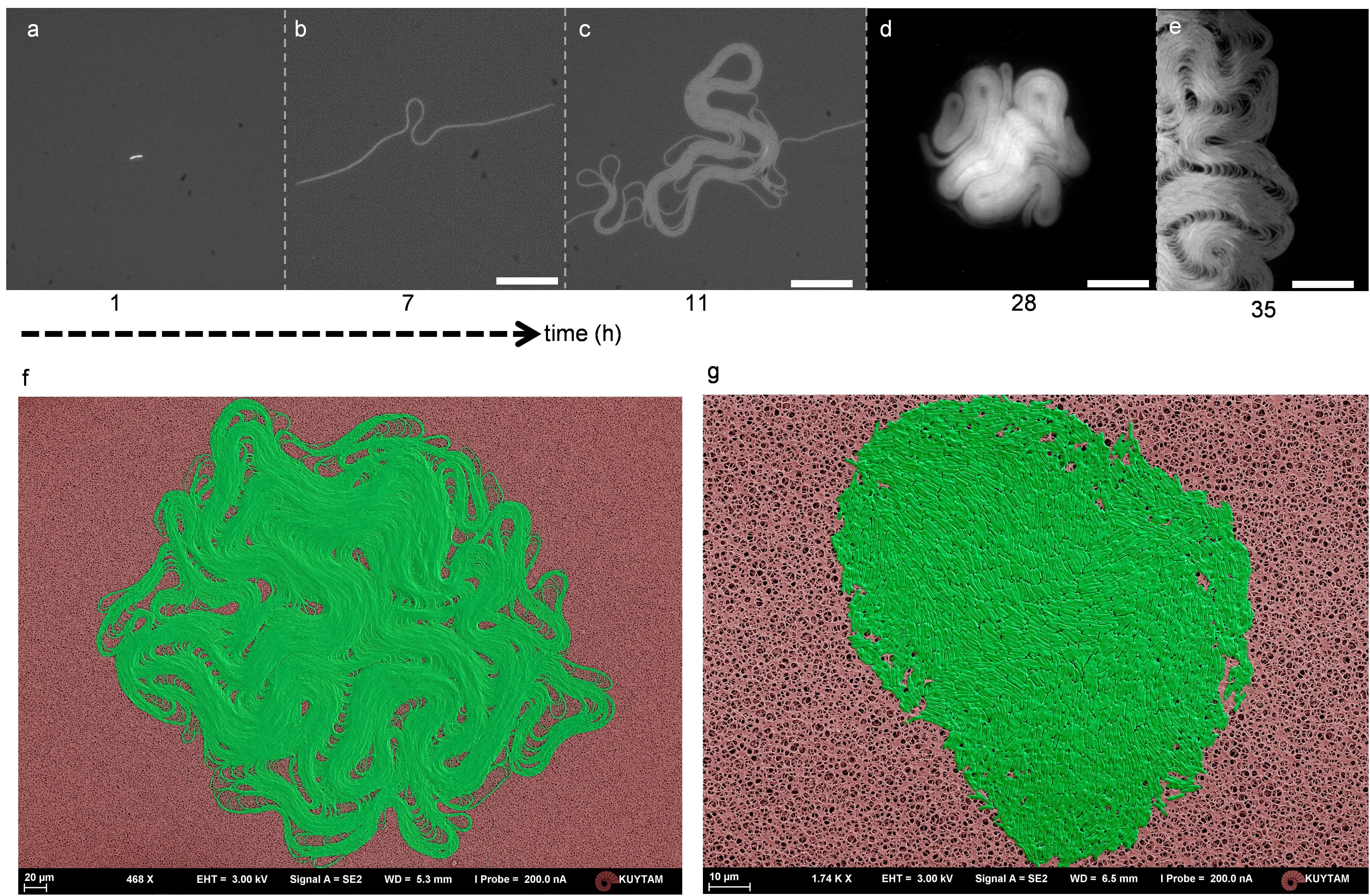}
		\caption{\textbf{Emergence of active nematics in a bacterial biofilm.} Snapshots of a growing biofilm. \textbf{a} The bacterium initially elongates and forms a bacterial chain through proliferation ($t = 2.5 h$). Scale bar, $30 \mu m$. \textbf{b} Due to local stress, the chain of attached bacteria buckles ($t = 7 h$). Scale bar, $30 \mu m$. \textbf{c} Multilayered and circular structures appear at the region where the localized buckling takes place ($t = 11 h$). Scale bar, $30 \mu m$. \textbf{d} The biofilm formation ($t = 28 h$). Scale bar, $150 \mu m$. \textbf{e} Kink bands appear and separate different domains across the biofilm. Scale bar, $30 \mu m$. Scanning electron microscopy images of \textbf{f} the biofilm formed by GFP labeled laboratory strain 168 (BAK47) and \textbf{g} the isolated bacterial colony formed by the non-chaining strain. All colonies were grown at T = 21 $^o$C.}
\end{figure*}

We first used time-lapse fluorescence microscopy to observe the temporal evolution of a biofilm formation. Figure 1 shows the snapshots from the growth of a chaining \textit{B. subtilis} biofilm, laboratory strain 168 (Fig. 1). This strain apparently forms a biofilm on a solid agar surface. Initially, divided cells give rise to a long and straight chain of attached bacteria. After several cell divisions, the first mechanical instability occurs (Supplementary Movie 1).

Interestingly, unlike Euler instability, this initial buckling is very localized.  As the bacterial chain grows, the buckled region forms a crumpled structure (Fig. 1c). Some parts of this structure become multilayered and circular. Similarly, these multilayered structures continue to grow radially, and they split into two (Supplementary Figure 1). Moreover, sharp walls appear across the biofilm, and these bands are observed as dark lines separating different domains in fluorescence images (Fig. 1d, e). All the basic dynamics and shapes described above were observed repeatedly across the biofilm and generate perfectly aligned cellular structures (Fig. 1f). This aligned structure and its dynamics strongly resemble the nematic active matter systems, particularly microtubule-based AN$^{27,28}$ (Supplementary Movie 2, 3, 4). 

To clarify the structural differences between nematic biofilm and isolated bacterial colony, we grew a similar but non-chaining \textit{B. subtilis} strain (BAK51, a derivative of 3610) under the same environmental conditions. This wild isolate failed to form chaining on an agar surface but can form a biofilm at a later stage stochastically (Supplementary Movie 5). In our bacterial strains, the flagella producing gene (\textit{hag}) was also mutated to eliminate the swimming induced motion. Detailed scanning electron microscopy images show tightly packed cells with fairly smooth colony edges (Fig. 1g, Supplementary Figure 2). 
\begin{figure*}[t]
\centering
		\includegraphics[width=\linewidth]{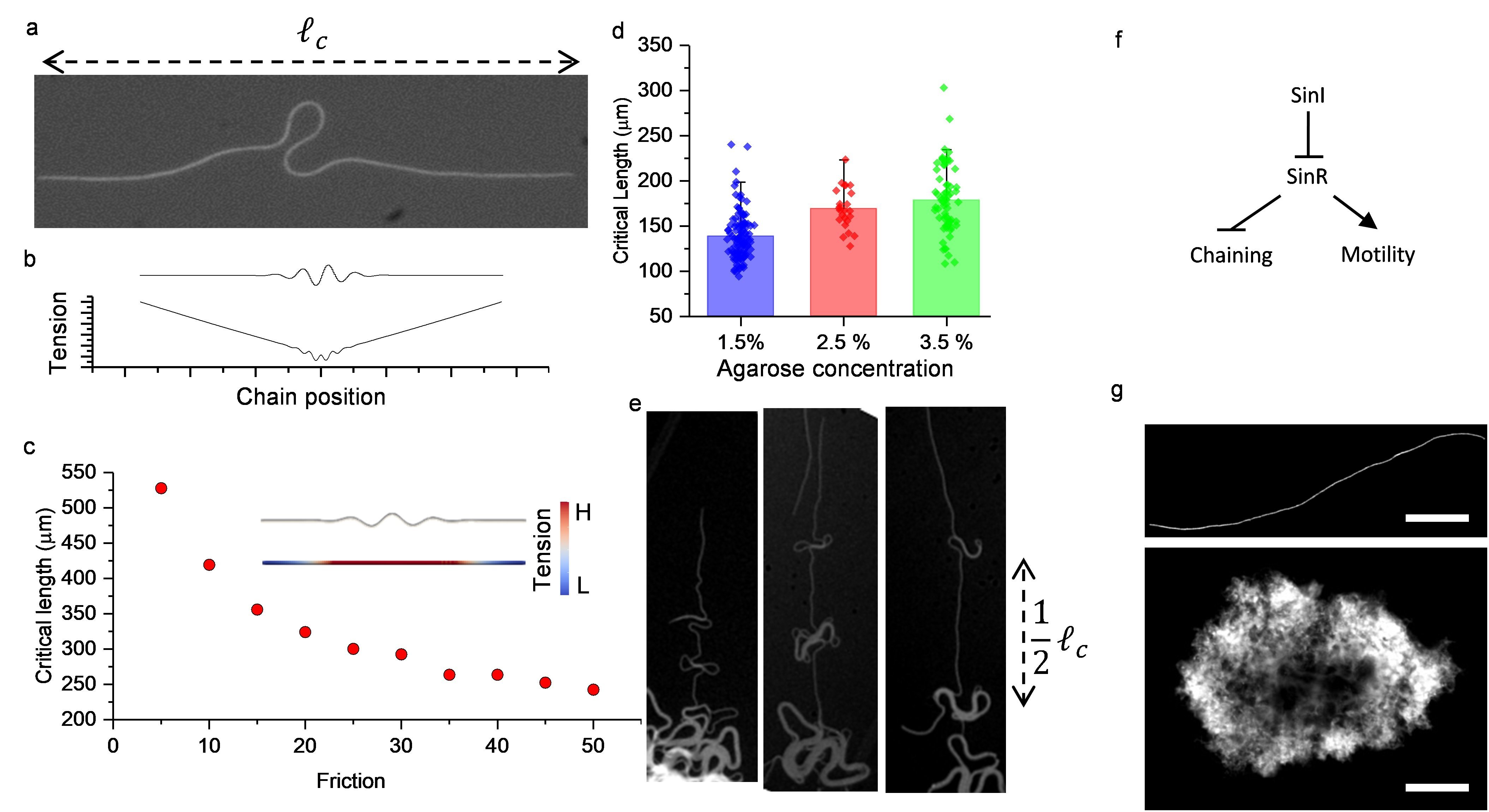}
		\caption{\textbf{Localized buckling of a bacterial chain}. \textbf{a} A snapshot of a bacterial chain with a length just above the critical buckling length. The buckled region is highly localized due to \textbf{b} localization of stress generated by the uniform friction force. The stress accumulates from the tip to the center of the chain due to outward axial velocity. The uniform friction force creates an almost linear stress distribution along the chain. \textbf{c} Plot of critical buckling length against friction, simulation results. Inlet shows a snapshot of the simulation just after the buckling and the color code shows the tension at pre-buckling stage. \textbf{d} Successive localized buckling instabilities on 1.5 \%, 2.5 \%, and 3.5 \%, respectively. After the first buckling, the subsequent buckling occurs at the middle of the straight part of the chain, which implies that the stress is symmetric around the center, and the connection of the former buckling region with the straight part is effectively a free end. Therefore, the separation between buckled regions is equal to half of the critical buckling length. \textbf{e} Experimental results of the critical buckling length for various agarose concentrations. The data is collected from more than 100 chains. \textbf{f} Schematic of a simplified regulatory network that controls the biofilm formation. The protein \textit{sinI} regulates the chaining state through double-negative feedback. \textbf{g} Bacterial chain grown in liquid LB. The chaining formation is initiated by IPTG induction. In the later stages, the colony forms a biofilm via supercoiling. Scale bar, $50 \mu m$ (top) and $200 \mu m$ (bottom).}
\end{figure*}

Clear nematic alignment of the biofilm required optimization of several parameters. First, we reduced the growth rate by decreasing the ambient temperature to eliminate the twisting and breakage of a bacterial chain (Supplementary Figure 3). Second, GFP labeling is essential to observe finely aligned structures, but low power fluorescence excitation is also necessary to eliminate light toxicity during biofilm development. 

Localized buckling appears to be the first building block of the nematic biofilm (Fig. 2a). To explore the physical mechanism underlying localization, we quantified the buckling condition. The localized buckling occurs just after reaching the critical chain length. As the chain continues to grow,  extended arms follow the same localized buckling scenario.  

The elasticity theory explaining the localization of the buckling is surprisingly complex. Both linear and nonlinear elastic properties of soft materials can contribute to the localization and formation of the post-buckling shape$^{29-34}$. However, the theoretical framework for growth-induced localized buckling is not well-studied. We have noticed that railroad thermal buckling$^{35}$ and growing plant roots$^{36}$ show a similar spatial localization profile. Therefore, we used the notation and the framework developed for these systems and followed the slender-body approximation to simulate the dynamics of a growing bacterial chain. We ignored the twisting dynamics, and we modeled the bacterial chain as an incompressible flexible elastic rod. Since the system is at low Reynolds number regime, we neglected the inertial forces and used the equations for mechanical equilibrium. According to the slender-body theory, in mechanical equilibrium, the following equation is satisfied.
\begin{equation}
\frac{d\textbf{F}(s)}{ds}=-\textbf{K}(s)
\end{equation} 
Where $\textbf{F}(s)$ is the internal force, and $\textbf{K}(s)$ is the external force acting on a unit length of the rod. We are interested in the pre-buckling tension profile. Therefore, we assumed that the bacterial chain is a straight rod with total length $\ell$ lying on $x$-axis. Hence the tension can be written as:
\begin{equation}
T(x)=f(abs(x)-\ell/2)
\end{equation}
where $f$ is the uniform friction force acting on a unit length of the rod (see Supplementary Information 1 for a detailed derivation). $T$ is linear and always negative implying that the rod is compressed. Tension profile locally exceeds the critical buckling limits when the total length reaches the threshold (Fig. 2b). This buckling condition sets a length scale for a bacterial chain. To further quantify the buckling conditions, we performed FEM simulations based on recently developed algorithms used to study growing elastic structures$^{37-39}$. Using realistic bacterial parameters, we found that an increase in friction force reduces the critical length (Fig. 2c). To experimentally verify this effect, we tuned the agarose concentration which serves as a flat surface and imaged the periodically buckled regions of a single bacterial chain. Our experiments revealed that the critical length changes with the agarose concentration (Fig. 2d, e). Typically, the bacterial chain becomes unstable above $120\mu m$. Interestingly, high agarose concentration reduces the friction force and increases the critical length. The exact mechanism behind the friction$^{40}$ is not clear to us; we speculate that the soft agar surface allows more local deformations and increases the friction force.  

Further, we tested the local buckling in a liquid environment which eliminates the friction force. In its wild-type form, \textit{B. subtilis} suppress biofilm formation in liquid LB culture. Detailed molecular genetics studies have identified that protein \textit{sinI} initiates and maintains the chaining state through double-negative feedback$^{1,3,4}$ (Fig. 2f). To trigger biofilm formation in liquid, protein \textit{sinI} should be overexpressed. To overcome this limitation, we used a genetically modified strain to drive \textit{sinI} using IPTG based induction (the background strains TMN1152 were received from R. Losick Lab). When we grew the biofilm in liquid culture with IPTG, local buckling disappeared, but the supercoiling process dominated the formation of the biofilm (Fig. 2g. Supplementary Movie 6, 7). We further clarified the structure of supercoiled bundles using fluorescence time-lapse microscopy and SEM imaging (Supplementary Figure 4). Altogether, we showed that friction force between the agar surface and the bacterial chain controls the mechanical instability. Localization of stress further generates spatially localized buckling and defines the critical length scale for a biofilm.

\begin{figure}[h]
\centering
		\includegraphics[width=1\textwidth]{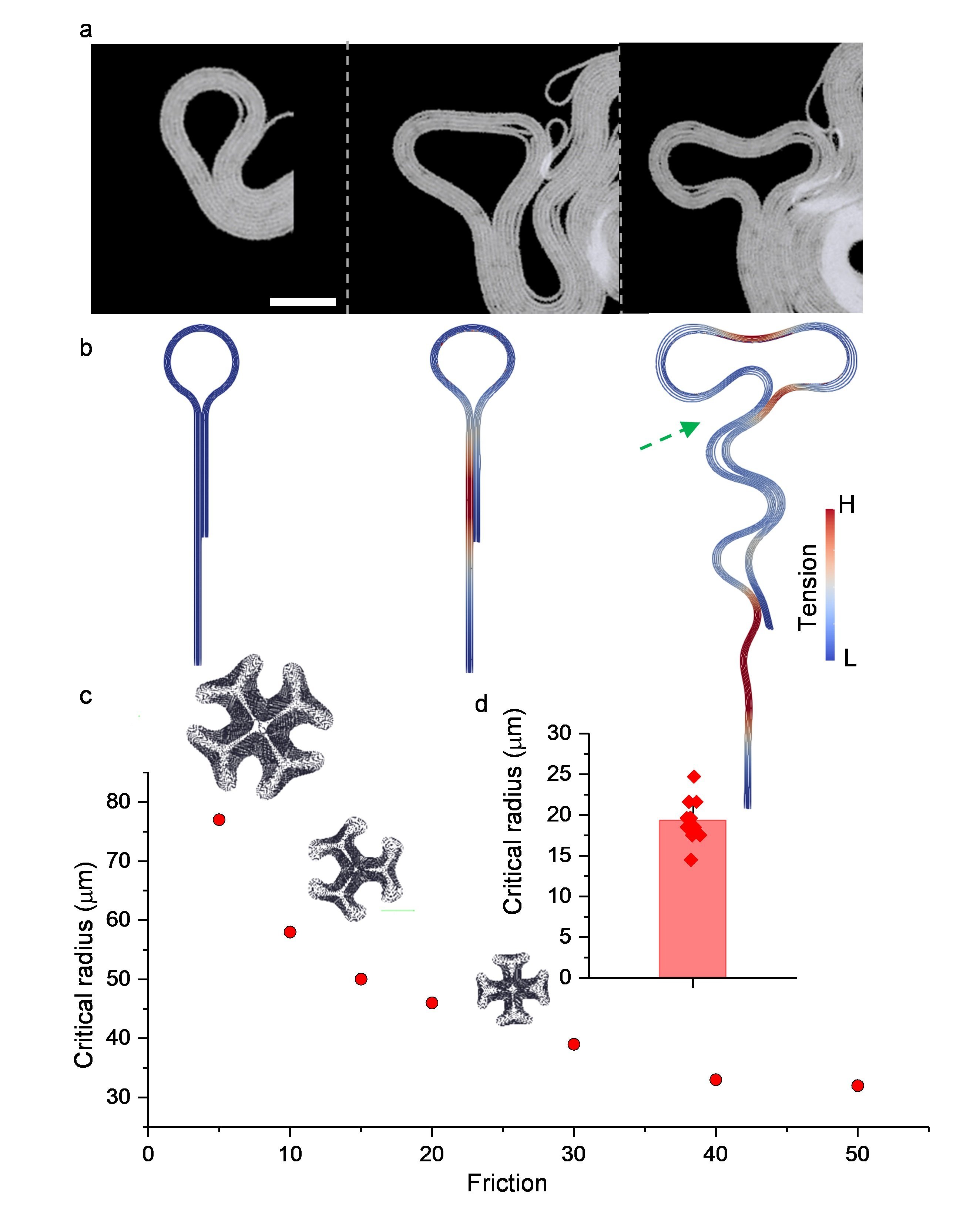}
		\caption{\textbf{Edge instabilities of a growing biofilm.} \textbf{a} Snapshots of growing multilayered circles. The radius reaches the critical radius and splits into two. Scale bar, 30$\mu m$ \textbf{b} Snapshots from the simulation of a droplet-like structure with a long tail. Color code shows the stress. While the droplet splits into two circles, the long tail locally buckles since its length is above the critical buckling length. \textbf{c} Multilayered growing circles are simulated under various friction forces. The plot of critical radius against friction, simulation results. Units are arbitrary. All snapshots were taken as the secondary edge instability occurred. \textbf{d} Experimental data of the critical radius. The colonies were grown on 1.5 \% agarose concentration.}
\end{figure}
As a second step, we focused on the edge instabilities of a growing biofilm. The growing multilayered circles are the most distinct geometric features observed around the leading edge (Fig. 3a). At first glance, the dynamics of these structures resemble the fingering instabilities of Hele-Shaw cell$^{41}$ and smectic-A liquid crystal filaments$^{42,43}$. These circles are connected to the film with a tail, and they resemble growing droplets. As they grow outward, instability occurs, and the droplets split into two. As gleaned from previous studies, we tested whether these structures have a characteristic critical radius which triggers new mechanical instabilities. We performed FEM simulations to observe the splitting dynamics. First, we simulated the growing multilayered circular structure with a tail (Fig. 3b, Supplementary Movie 8). We found that these structures indeed have a critical radius. Above this radius, circular structures are unstable and split into smaller but stably growing ones. As we observed in previous experiments, the straight tail also locally buckles in the direction perpendicular to growing axis after reaching the critical length. Our simulation verified that edge instability occurs when the circumference of the circular structure exceeds the linear critical length $2\pi R_c=\ell_c$ (Fig. 3c, d). We confirmed this relationship experimentally. Similarly, both a critical radius and a critical length are controlled by friction.

We also observed that, just after the splitting, an additional buckling might occur around the junction point (connection between the droplet and the tail) where the chain has the highest curvature (green arrow in Fig. 3b). This buckling further makes the divided droplet  fall back on the biofilm edge asymmetrically (Supplementary Figure 5). We then extended our simulations to observe large-scale biofilm growth starting from many concentric circular rods (Supplementary Movie 9). As expected, friction shapes the overall biofilm morphology by defining the maximum radius of the curvature around the leading edges (Fig. 3d). Altogether our results show that a growing biofilm has a critical radius which controls the dynamics of edge instabilities. 

One of the characteristics of active nematics is the formation of topological defects$^{16-18,27,28,44}$. To better understand the active matter nature of a biofilm, we further focused on how mechanical instabilities generate topological defects. In microtubule-based active matter systems, topological defects (+1/2 or -1/2) (Fig. 4a) are spontaneously created and annihilated. The exact mechanism behind the creation of these defects remains unknown$^{45}$. Recent studies elegantly demonstrate that the confinement strongly controls defects formation$^{45,46}$. In contrast, our biofilm system does not require any physical confinement which enables the edge to freely move and allows observation of the evolution of the defect formation. Analyzing the development progress, we found two main defect forming mechanisms; the large-scale biofilm folding and edge instabilities.  
  \begin{figure*}
\centering
		\includegraphics[width=\linewidth]{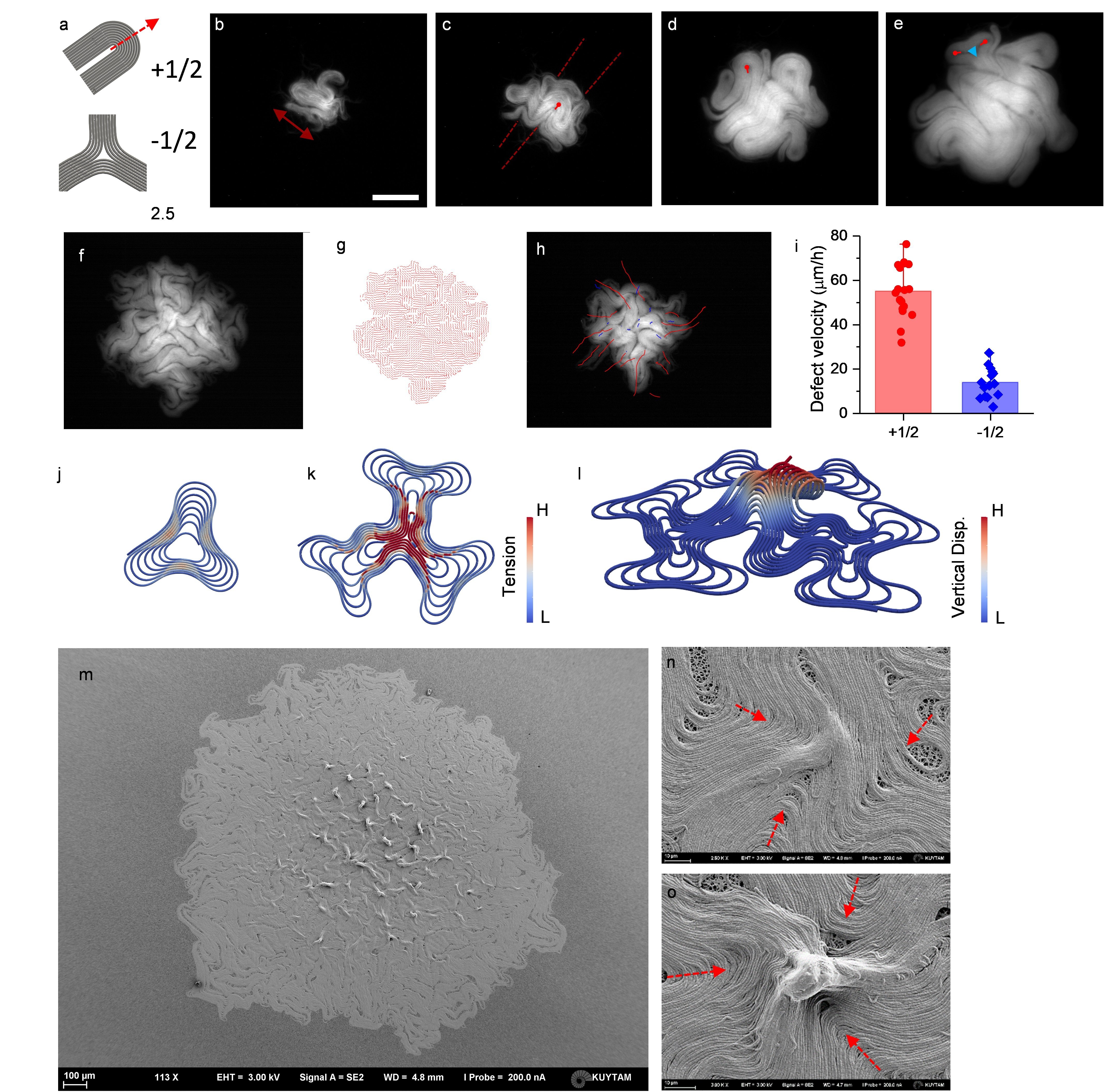}
		\caption{\protect\rule{0ex}{0ex}\textbf{Dynamics of topological defects and vertical lift-off.} \textbf{a} Comet-like motile (+1/2) and trefoil-like (-1/2) topological defects. \textbf{b}, \textbf{c}, \textbf{d}, \textbf{e} Snapshots of a growing biofilm. \textbf{b} Biofilm elongates and reaches the critical buckling length. Scale bar, 100$\mu m$. \textbf{c} As a result, biofilm shows a structural folding and creates (+1/2) topological defects. \textbf{d}, \textbf{e} (+1/2) topological defects at the edge move outward, split into two (+1/2) defects and create a (-1/2) defect at the junction. \textbf{f} Snapshot of a growing biofilm. \textbf{g} The director field of the biofilm (\textbf{f}). \textbf{h} Trajectories of both (+1/2) and (-1/2) defects. Red lines indicate the trajectories of (+1/2) defects which are motile, whereas blue lines represent the trajectories of (-1/2) defects, which are stationary. \textbf{i} Experimental data of velocity distribution of (+1/2) and (-1/2) defects.  \textbf{j}, \textbf{k}, \textbf{l} Simulation results of a growing multilayered circular colony. \textbf{j} After reaching the critical radius, the colony folds and creates topological defects. \textbf{k} Inward propagating defects collide and create a (-1/2) defect. The stress is accumulated at the collision point, namely (-1/2) topological defect. \textbf{l} The accumulated stress results in vertical lifting around the (-1/2) topological defect. \textbf{m}, \textbf{n}, \textbf{o} SEM images of a biofilm in late stage. Red arrows indicate the propagation directions of defects. \textbf{m} Vertically lifted domains are in the vicinity of the center where the stress accumulates. \textbf{n} Motile (+1/2) defects collide and form a stationary (-1/2) defect. \textbf{o} Accumulation of stress causes vertical lift-off.
}
\end{figure*}

Using time-lapse microscopy, we recorded the first defect generation process. As it grows, the entire biofilm reaches the critical length scale and triggers large structural folding (Fig. 4b, c). This folding clearly forms  +1/2  defects. Fig. 4 shows the temporal evolution of this defect formation. As observed, the biofilm sequentially folds in orthogonal directions and generates several motile +1/2 defects (Supplementary Figure 6). Clearly, in this stage the total number of defects increases. Due to the small size, all defects move outward. As the size of a biofilm increases, the critical length drops below the biofilm size, and this type of folding appears locally (Supplementary Movie 2, 3).

We also found that the edge instabilities is the second type of defect forming mechanism. Growing droplet geometries around the edge simply behave like +1/2 topological defects (Fig. 4d). These structures are very motile because the unbalanced force generated by the growing side chains can push the defect forward. In contrast to the structural folding described above, this growing droplet splits into two and generates paired +1/2 defects, and it leaves one -1/2 defect around the junction point (Fig. 4e). This mechanism keeps the total topological charge constant . At later stages foldings and splittings become strongly coupled. Thus, the dynamics of defect formation become very chaotic.

Our simulations verified the dynamics of defect formation around the edge . We observed that above the critical radius both outward and inward propagating bucklings appear. Accumulated localized stress gradually slow down the speed of the central region and then starts migrating inward. These inward propagating chains might collide with the junction point and then become a stationary  -1/2 defect by leaving two motile paired +1/2 (Supplementary Movie 9). 

Notably, some of the propagating defects leave king bands as a clear trace (Fig. 4f). These bands originate from the differences in the radius of multi-layered structures which eventually define the total growth rate of the circumference. 

To further examine defect formations experimentally, we measured the speed distributions of defects within a biofilm (Fig. 4i). We found that +1/2 defects are more motile around the edge and -1/2 defects are stationary.  We also noticed that as the biofilm develops the central region becomes more stationary. The most parsimonious hypothesis based on these observations is that stress accumulates around the center and slows down the dynamics.

Topological defects could also change the physiology of the biological tissues$^{18,47-49}$. It was shown that defect-induced stress triggers mechanotransduction and alters the genes expressions$^{16}$. Stress also modulates the cellular density of cultured cells. Topological defects and active stress could also play a biophysical role during the development of a biofilm. With this motivation, we further performed stress analysis of topological defects. Our simulations revealed that the stress accumulates as stationary -1/2 defect occurs (Fig. 4j, k). Around these defects, the unbalanced force generated by the growing chain can be balanced by counter propagating ones. Further, colliding bent chains significantly increase the total stress. When we performed our simulation in the three-dimensional domain (so far we performed all simulation in a 2D confined environment), we observed vertical lifting of the elastic rods around the vicinity of -1/2 defect (Fig. 4l, Supplementary Movie 10). 

Finally, we asked whether we could experimentally observe the effects of stress accumulation in a biofilm.  To do so, we analyzed the 3D surface topology after getting a stable biofilm (35 h after seeding). SEM images revealed vertical structures around the -1/2 defects sites (Fig. 4m, n, o). A close investigation clarified that these structures are actually the vertically lifted parts of the defects. These structures are more visible around the center of the biofilm. We grew the non-chaining bacteria in the same conditions and we did not observe any of these vertical structures (Supplementary Figure 7). The results are remarkably consistent with our FEM simulations. These structures have already been identified as aerosol or fruiting bodies of the biofilm which eventually produce bacterial spores$^{2}$. We do not exclude other structural instabilities or the contribution of twisting process. Our results support the idea that topological defects are one of the possible origins of these vertical ordered geometries.
\section*{Discussion}
The theory of active nematics provides a robust framework to explain the coordinated motion of cells and the emergence of large-scale order in biological systems. Our study throws light upon different aspects of biological active matter systems: First, mechanical instabilities can be simply characterized by a critical length scale. Different types of length scales have previously been defined for various AN systems$^{24,26}$. These quantities mainly describe the activity of the system. In contrast, our length scale reflects the elastic properties of the biofilm. Second, the leading edge of the biofilm is not stable. Edge instabilities$^{15,44,50-52}$ have received significant attention due to fingering formation during wound healing$^{15}$. Our results revealed the importance of cellular alignment and localized stress driving instabilities. Finally, we have shown the details of defect-forming mechanisms and their effect on the stress distribution across the biofilm. In growing tissue or bacterial colonies, mechanical stress can easily result in physiological stress. Recent studies have linked the physical forces with the cellular response$^{16}$, and our system could further expand the understanding of stress management across growing tissue$^{53}$. Another benefit of using \textit{B. subtilis} is the availability of sophisticated genetic toolbox$^{54}$. Genetic manipulations and control could reveal unexplored dynamics of active matter systems. We believe, our biofilm platform offers new opportunities to study active nematics in living systems.  
\vspace{-\baselineskip}
\subsection*{Methods}

\begin{footnotesize}
\textbf{Sample preparation and growth conditions.} Bacterial cultures (BAK47 and BAK51) were grown in Luria-Bertani (LB) broth at 37$^o$C on a shaker. An overnight culture was diluted 100x and grown 8 h. The culture was diluted 10000x and 50$\mu \ell$ of culture was seeded on an LB agarose plate. These isolated bacteria on plates were grown at 21$^o$C for 12 h then imaged. Cover glass was not used. IPTG (isopropyl $\beta$-D-1-thiogalactopyranoside) inducible (BAK50) were grown in LB broth at 37$^o$C without IPTG induction to ensure that chaining does not occur. An overnight culture was diluted 10000x and supplemented with 100$\mu M$ IPTG. A droplet of IPTG inducable bacterial suspension is placed between two glass slides with 100$\mu m$ separation, then imaged at 21$^o$C.

%\vspace{\baselineskip}
\noindent
\textbf{Microscopy Imaging.} Fluorescence time-lapse imaging was performed using a Nikon SMZ18 stereo microscope and images were obtained using a Thor Labs DCC1545M CMOS camera. Time intervals between successive images are 15 minutes. To obtain the best image, the light exposure is controlled adaptively for different colony sizes by a custom-written program in LabVIEW. A typical colony growth experiment was run for 1 day.

%\vspace{\baselineskip}
\noindent
\textbf{SEM imaging.} Before seeding the bacteria on LB agar surface, sterile filter paper with pore size 0.2 $\mu m$ is placed on the surface. After seeding the bacteria on the filter paper, bacteria were grown on the paper for 24 h at 21$^o$C. Then the filter paper was peeled off from the surface and the colonies were fixed using paraformaldehyde and left to drying. Fixed colonies were imaged using Zeiss Ultra Plus Field Emission Electron Microscope. 

%\vspace{\baselineskip}
\noindent
\textbf{Strains and labeling.}
\vspace{-3ex}
\begin{center}
\resizebox{\linewidth}{!}{
    \begin{tabular}{ | l | l | l | l |}
    \hline
    Strain & Parent & Operation & Genotype\\ \hline
    BAK47 & 168 & Transformed with plasmid ECE321 & \textit{amyE}::$\text{P}_{\text{veg}}$-sfGFP (Spc)\\&&from Bacillus Genetic Stock Center&\\ \hline
    BAK50 & TMN1152 & Transformed with plasmid ECE321 & \textit{amyE}::$\text{P}_{\text{veg}}$-sfGFP (Spc)\\&&from Bacillus Genetic Stock Center& \textit{ywrK}::$\text{P}_{\text{spank}}$-\textit{sinI} (Spc)\\ \hline
    BAK51 & TMN1138 & Transformed with plasmid ECE321& \textit{amyE}::$\text{P}_{\text{veg}}$-sfGFP (Spc) \\&&from Bacillus Genetic Stock Center& \textit{sacA}::$\text{P}_{\text{hag}}$-mKate2L (Kan)\\&&&\textit{hagA233V} (Phleo)\\ \hline 
    \end{tabular}}
\end{center}

%\vspace{\baselineskip}
\noindent
\textbf{Detection of topological defects.} Images were smoothed using Bandpass filter in ImageJ, and Coherence-Enhanced Diffusion Filter was applied to images in MATLAB. The ImageJ plugin OrientationJ was used to find the nematic director field of the colony. We followed the defect detection algorithm$^{27}$ using a custom-written MATLAB code.

%\vspace{\baselineskip}
\begin{spacing}{1.03}
\noindent
\textbf{Computer Simulations.} For the computer simulations involving large growth, we employed a dynamic finite element program$^{37-39}$. The filament was modelled as an isotropic, linearly elastic continuum with three-dimensional beam theory using cubic Hermite shape functions and a corotational quaternion formulation for geometric nonlinearity. The filament was assumed to have a circular cross section with constant uniform radius $r=0.7\mu m$, a Young's modulus of $E=5300Pa$ (corresponding to a bending stiffness of $E\pi r^4/4={10}^{-21}Nm^2)$, a Poisson ratio of 0.33 and a mass density of $1g/cm^3$. The elastic energy of the filament consisted of the sum of bending, torsion and axial compression energies. Hertzian repelling forces were exchanged between contacting filament elements in normal direction, whereas the tangential contact forces were realized with a Coulomb slip-stick friction model. Newton's equations of motion were integrated in time with a second-order predictor-corrector scheme. For numerical robustness and to keep the filament near static equilibrium during growth, subcritical damping forces were added. The initial configuration was assumed to be stress-free, and the filament was grown exponentially in length over time, according to $\ell(t)=\ell_0e^{\sigma t}$ where $\sigma$ is the exponential growth rate. This was realized by continuously increasing the equilibrium length of each rod element. The substrate was modeled as an ideal elastic horizontal plane onto which the filament was placed, and a gravitational force was applied perpendicular to the plane.
\end{spacing}
\end{footnotesize}
\vspace{-\baselineskip}
\subsubsection*{Acknowledgements}
\begin{footnotesize}
\begin{spacing}{1.03}
\vspace{-2ex}
This work was supported by an EMBO installation Grant (IG 3275, A.K.) and BAGEP young investigator award (A.K.).
\end{spacing}
\end{footnotesize}
\vspace{-\baselineskip}
\subsubsection*{Author Contributions}
\begin{footnotesize}
\begin{spacing}{1.03}
\vspace{-2ex}
Y.I.Y and A.K. designed and performed experiments, analyzed data and developed the imaging systems. Y.I.Y and E.D. A.K. performed the molecular biology experiments. R.V. designed and developed the FEM simulation toolbox and optimized the codes with realistic bacterial parameters. Y.I.Y performed the simulations. Y.I.Y and A.K. prepared the draft, and all authors contributed to the writing of the manuscript.
\end{spacing}
\end{footnotesize}
\vspace{-\baselineskip}
\subsubsection*{Competing Interests}
\begin{footnotesize}
\vspace{-2ex}
Authors declare no competing interests.
\end{footnotesize}
\vspace{-\baselineskip}
\subsection*{References}
\begin{footnotesize}
\begin{enumerate}
\item	Chai, Y. R., Norman, T., Kolter, R. \& Losick, R. An epigenetic switch governing daughter cell separation in Bacillus subtilis. \textit{Gene Dev} \textbf{24}, 754-765 (2010).

\item	Branda, S. S., Gonzalez-Pastor, J. E., Ben-Yehuda, S., Losick, R. \& Kolter, R. Fruiting body formation by Bacillus subtilis. \textit{P Natl Acad Sci USA} \textbf{98}, 11621-11626 (2001).

\item	Norman, T. M., Lord, N. D., Paulsson, J. \& Losick, R. Memory and modularity in cell-fate decision making. \textit{Nature} \textbf{503}, 481-+ (2013).

\item	Chai, Y. R., Kolter, R. \& Losick, R. Reversal of an epigenetic switch governing cell chaining in Bacillus subtilis by protein instability. \textit{Mol Microbiol} \textbf{78}, 218-229 (2010).

\item	Vlamakis, H., Chai, Y. R., Beauregard, P., Losick, R. \& Kolter, R. Sticking together: building a biofilm the Bacillus subtilis way. \textit{Nature Reviews Microbiology} \textbf{11}, 157-168 (2013).

\item	Liu, J. T. et al. Coupling between distant biofilms and emergence of nutrient time-sharing. \textit{Science} \textbf{356}, 638-641 (2017).

\item	Liu, J. T. et al. Metabolic co-dependence gives rise to collective oscillations within biofilms. \textit{Nature} \textbf{523}, 550-+ (2015).

\item	Mamou, G., Mohan, G. B. M., Rouvinski, A., Rosenberg, A. \& Ben-Yehuda, S. Early Developmental Program Shapes Colony Morphology in Bacteria. \textit{Cell Reports} \textbf{14}, 1850-1857 (2016).

\item	Honda, R., Wakita, J. I. \& Katori, M. Self-Elongation with Sequential Folding of a Filament of Bacterial Cells. \textit{Journal of the Physical Society of Japan} \textbf{84} (2015).

\item	Klapper, I. Biological applications of the dynamics of twisted elastic rods. \textit{J Comput Phys} \textbf{125}, 325-337 (1996).

\item	Wolgemuth, C. W., Goldstein, R. E. \& Powers, T. R. Dynamic supercoiling bifurcations of growing elastic filaments. \textit{Physica D} \textbf{190}, 266-289 (2004).

\item	Mendelson, N. H. Bacillus subtilis macrofibres, colonies and bioconvection patterns use different strategies to achieve multicellular organization. \textit{Environ Microbiol} \textbf{1}, 471-477 (1999).

\item	Mendelson, N. H., Thwaites, J. J., Kessler, J. O. \& Li, C. Mechanics of Bacterial Macrofiber Initiation. \textit{J Bacteriol} \textbf{177}, 7060-7069 (1995).

\item	van Gestel, J., Vlamakis, H. \& Kolter, R. From Cell Differentiation to Cell Collectives: Bacillus subtilis Uses Division of Labor to Migrate. \textit{Plos Biol} \textbf{13} (2015).

\item	Basan, M., Elgeti, J., Hannezo, E., Rappel, W. J. \& Levine, H. Alignment of cellular motility forces with tissue flow as a mechanism for efficient wound healing. \textit{P Natl Acad Sci USA} \textbf{110}, 2452-2459 (2013).

\item	Saw, T. B. et al. Topological defects in epithelia govern cell death and extrusion. \textit{Nature} \textbf{544}, 212-+ (2017).

\item	Duclos, G., Erlenkamper, C., Joanny, J. F. \& Silberzan, P. Topological defects in confined populations of spindle-shaped cells. \textit{Nature Physics} \textbf{13}, 58-62 (2017).

\item	Kawaguchi, K., Kageyama, R. \& Sano, M. Topological defects control collective dynamics in neural progenitor cell cultures. \textit{Nature} \textbf{545}, 327-+ (2017).

\item	Doostmohammadi, A., Thampi, S. P. \& Yeomans, J. M. Defect-Mediated Morphologies in Growing Cell Colonies. \textit{Phys Rev Lett} \textbf{117} (2016).

\item	Dell'Arciprete, D. et al. A growing bacterial colony in two dimensions as an active nematic. \textit{Nat Commun} \textbf{9} (2018).

\item	Wensink, H. H. et al. Meso-scale turbulence in living fluids. \textit{P Natl Acad Sci USA} b, 14308-14313 (2012).

\item   Beroz, F. et al. Verticalization of bacterial biofilms. \textit{Nature Physics} \textbf{14}, 954-+, doi:10.1038/s41567-018-0170-4 (2018).

\item   Hartmann, R. et al. Emergence of three-dimensional order and structure in growing biofilms. \textit{Nature Physics}, doi:10.1038/s41567-018-0356-9 (2018).

\item	Guillamat, P., Ignes-Mullol, J. \& Sagues, F. Control of active liquid crystals with a magnetic field. \textit{P Natl Acad Sci USA} \textbf{113}, 5498-5502 (2016).

\item	Guillamat, P., Ignes-Mullol, J., Shankar, S., Marchetti, M. C. \& Sagues, F. Probing the shear viscosity of an active nematic film. \textit{Phys Rev E} \textbf{94} (2016).

\item	Guillamat, P., Ignes-Mullol, J. \& Sagues, F. Taming active turbulence with patterned soft interfaces. \textit{Nat Commun} \textbf{8} (2017).

\item	DeCamp, S. J., Redner, G. S., Baskaran, A., Hagan, M. F. \& Dogic, Z. Orientational order of motile defects in active nematics. \textit{Nat Mater} \textbf{14}, 1110-1115 (2015).

\item	Sanchez, T., Chen, D. T. N., DeCamp, S. J., Heymann, M. \& Dogic, Z. Spontaneous motion in hierarchically assembled active matter. \textit{Nature} \textbf{491}, 431-+ (2012).

\item	Hunt, G. W., Wadee, M. K. \& Shiacolas, N. Localized Elasticae for the Strut on the Linear Foundation. \textit{J Appl Mech-T Asme }\textbf{60}, 1033-1038 (1993).

\item	Shan, W. L. et al. Attenuated short wavelength buckling and force propagation in a biopolymer-reinforced rod. \textit{Soft Matter} \textbf{9}, 194-199 (2013).

\item	Diamant, H. \& Witten, T. A. Compression Induced Folding of a Sheet: An Integrable System. \textit{Phys Rev Lett} \textbf{107} (2011).

\item	Diamant, H. \& Witten, T. A. Shape and symmetry of a fluid-supported elastic sheet. \textit{Phys Rev E} \textbf{88} (2013).

\item	Oshri, O., Brau, F. \& Diamant, H. Wrinkles and folds in a fluid-supported sheet of finite size. \textit{Phys Rev E} \textbf{91} (2015).

\item	Thompson, J. M. T. \& Champneys, A. R. From helix to localized writhing in the torsional post-buckling of elastic rods. \textit{P Roy Soc a-Math Phy} \textbf{452}, 117-138 (1996).

\item	Tvergaard, V. \& Needleman, A. On Localized Thermal Track Buckling. \textit{Int J Mech Sci} \textbf{23}, 577-587 (1981).

\item	Silverberg, J. L. et al. 3D imaging and mechanical modeling of helical buckling in Medicago truncatula plant roots. \textit{P Natl Acad Sci USA} \textbf{109}, 16794-16799 (2012).

\item	Vetter, R., Wittel, F. K., Stoop, N. \& Herrmann, H. J. Finite element simulation of dense wire packings. \textit{Eur J Mech a-Solid} \textbf{37}, 160-171 (2013).

\item	Vetter, R., Wittel, F. K. \& Herrmann, H. J. Morphogenesis of filaments growing in flexible confinements. \textit{Nat Commun} \textbf{5} (2014).

\item	Vetter, R. Growth, Interaction and Packing of Thin Objects, \textit{ETH Zurich}, (2015).

\item	Grant, M. A. A., Waclaw, B., Allen, R. J. \& Cicuta, P. The role of mechanical forces in the planar-to-bulk transition in growing Escherichia coli microcolonies. \textit{J R Soc Interface }\textbf{11} (2014).

\item	Paterson, L. Radial Fingering in a Hele Shaw Cell. \textit{J Fluid Mech} \textbf{113}, 513-529 (1981).

\item	Naito, H., Okuda, M. \& Zhongcan, O. Y. Pattern formation and instability of smectic-A filaments grown from an isotropic phase. \textit{Phys Rev E} \textbf{55}, 1655-1659 (1997).

\item	Shelley, M. J. \& Ueda, T. The Stokesian hydrodynamics of flexing, stretching filaments. \textit{Physica D} \textbf{146}, 221-245 (2000).

\item	Blow, M. L., Thampi, S. P. \& Yeomans, J. M. Biphasic, Lyotropic, Active Nematics. \textit{Phys Rev Lett} \textbf{113} (2014).

\item	Opathalage, A., Norton, M. M., Juniper, M. P. N., Aghvami, S. A., Langeslay, B., Fraden, S. \& Dogic, Z. (arXiv:1810.09032, 2018).

\item	Norton, M. M. et al. Insensitivity of active nematic liquid crystal dynamics to topological constraints. \textit{Phys Rev E} \textbf{97} (2018).

\item	Hawkins, R. J. \& April, E. W. Liquid-Crystals in Living Tissues. \textit{Adv Liq Cryst} \textbf{6}, 243-264 (1983).

\item	Bonhoeffer, T. \& Grinvald, A. Iso-Orientation Domains in Cat Visual-Cortex Are Arranged in Pinwheel-Like Patterns. \textit{Nature} \textbf{353}, 429-431 (1991).

\item	Gruler, H., Dewald, U. \& Eberhardt, M. Nematic liquid crystals formed by living amoeboid cells. \textit{Eur Phys J B} \textbf{11}, 187-192 (1999).

\item	Nesbitt, D., Pruessner, G. \& Lee, C. F. Edge instability in incompressible planar active fluids. \textit{Phys Rev E} \textbf{96} (2017).

\item	Zimmermann, J., Basan, M. \& Levine, H. An instability at the edge of a tissue of collectively migrating cells can lead to finger formation during wound healing. \textit{Eur Phys J-Spec Top} \textbf{223}, 1259-1264 (2014).

\item	Doostmohammadi, A. et al. Celebrating Soft Matter's 10th Anniversary: Cell division: a source of active stress in cellular monolayers. \textit{Soft Matter} \textbf{11}, 7328-7336 (2015).

\item	Martinez-Corral, R., Liu, J. T., Suel, G. M. \& Garcia-Ojalvo, J. Bistable emergence of oscillations in growing Bacillus subtilis biofilms. \textit{P Natl Acad Sci USA} \textbf{115}, E8333-E8340 (2018).

\item	Guiziou, S. et al. A part toolbox to tune genetic expression in Bacillus subtilis. \textit{Nucleic Acids Res} \textbf{44}, 7495-7508 (2016).
\end{enumerate}
\end{footnotesize}
\end{document}